# Assessing the Effect of Coulomb Repulsion Asymmetry on Electron Pairing

*Farshid Raissi*

K. N. Toosi University of Technology, EE Dept.

P.O.BOX 16315-1355

Tehran 16314, Iran

Phone: +98-912-3161462, FAX: +98-21-8602066

raissi@kntu.ac.ir

ABSTRACT__ Coulomb repulsion between two moving electrons loses its spherical symmetry due to relativistic effects. In presence of a uniform positive ion background this asymmetry uncovers an angular dependent attraction potential in the direction of motion. The quantum mechanical response to such an attraction potential is obtained through perturbation. It is shown that the transition amplitude between sates with the symmetry of the attraction potential becomes negative and if the density of states is anisotropic, occurrence of a superconducting state becomes possible.

INTRODUCTION

Occurrence of superconducting state is due to the existence of an attraction force between electrons which can lead into a bound state of lower energy. In BCS theory the origin of this attraction is the electron-phonon-electron interactions. Low Tc superconductivity occurs when this attraction becomes larger than screened Coulomb repulsion. The transition amplitude between spherically symmetric wave functions (S orbital) due to the spherically symmetric Coulomb repulsion potential and phonon scattering is calculated and the energy lowering of the superconducting sate is obtained [1]. In this report we consider the effect of asymmetric Coulomb repulsion potential on scattering amplitude. We first look at a purely classical picture taking electron and nuclei as point charges with symmetric Coulomb potentials. Based on the principles of special relativity there is a marked difference between the stationary and moving potential of charged particles. If electrons are in motion their Coulomb potential changes such that a directional dependent attraction force is uncovered. Such a phenomenon might (and probably should) have a counterpart in quantum mechanics. We treat this new potential as a perturbation and will observe that the scattering amplitude between states with the symmetry of the attraction potential (mainly p and d-type states) becomes negative. It is shown that if the density of states is anisotropic and wave functions involved in scattering have p or d-type symmetry the total energy of the system is decreased and occurrence of a superconducting state becomes possible.

RELATIVISTIC COULOMB REPULSION POTENTIAL:

In the theory of relativity when an electron is in motion the repulsion potential loses its spherical symmetry [2]. It increases in directions perpendicular to its direction of motion and it decreases in directions parallel to it. The relativistic Coulomb repulsion is given below:

$$\Phi_R = \frac{q^2}{r} \frac{1 - \frac{v^2}{c^2}}{\left[1 - \frac{v^2}{c^2}\sin^2\theta\right]^{\frac{3}{2}}}, \tag{1}$$

where, $q$ is the electron charge, $r$ denotes the distance from moving electron, $v$ is the speed of the electron, $c$ is the speed of light in the medium and $\theta$ denotes the angle from the direction of motion.

It is worth mentioning that the so called magnetic attraction between two moving electrons is different from the subject of this paper. By principles of special relativity the repulsion force between two moving electrons decreases by a certain amount which in electromagnetic theory is attributed to magnetic force. Due to this relativistic effect the repulsion force at most decreases to zero at the speed of light. There is never a resultant attraction force between two moving electrons which are inside the same medium. When these electrons are moving inside different materials, such as inside two separate wires, an attraction force is created between the wires (due to length contraction and the change in total charge density). In this paper we are concerned with a different relativistic effect which is the change in electric potential around a moving electron as observed by positive stationary nuclei. And we will show that this relativistic effect can result in an attraction force between electrons. Equation 1 has not been included so far in any relativistic treatment of quantum mechanics and prior treatments have ignored its effect. The spin-orbit interaction, which is also a relativistic effect is different from this term as well. The relativistic repulsion potential is depicted in Fig.1 for an electron speed of 0.7 $c$. The stationary repulsion potential is also shown for comparison. As is observed the repulsion potential has decreased to 50% of its stationary value in the direction of motion.

In a classical treatment which ignores the wave nature of electrons, let us consider a crystal in which electrons are superimposed on a background of positive charges and the crystal is charge neutral. The electrostatic potential $\Phi$ is zero as found from the Poisson's equation $\nabla^2 \Phi = -4\pi\rho$, where $\rho$ is the charge density which is zero inside any closed volume. We have schematically drawn this situation in Fig. 2. There are positive ion cores drawn in a line whose spherically symmetric attraction potential $\Phi_{Attraction} = \frac{-q^2}{r}$ is compensated by the spherically symmetric repulsion potential $\Phi_{Re\,pulsion} = \frac{q^2}{r}$ due to electrons. The circles around each electron or positive core denote its respective potential. Electrons and ion cores have been misplaced with respect to each other for clarity otherwise they are superimposed on top of each other. As was mentioned above, the electrostatic potential $\Phi = \Phi_{Re\,pulsion} + \Phi_{Attraction}$ is zero everywhere.

Now if electrons are in motion their respective repulsion potential changes as given in equation 1. In Fig. 3 the new potentials are drawn separately along with the resultant net potential which is obtained by superimposing the stationary attraction potential of the positive ions by the relativistic repulsion force of the moving electrons. It is observed that in the direction of motion there exists a resultant attraction potential, and a repulsion potential has been created in directions perpendicular to the direction of motion. In other words the electrostatic potential $\Phi$ is no longer zero in all directions. (It must be kept in mind that still $\nabla^2\Phi = -4\pi\rho = 0$ when a closed volume is considered). This net nonzero electrostatic potential can be written as:

$$\Phi = -\frac{q^2}{r}(1 - \frac{1-\frac{v^2}{c^2}}{\left[1-\frac{v^2}{c^2}\sin^2\theta\right]^{\frac{3}{2}}}). \tag{2}$$

What this means is that as an electron speeds up an attraction force is developed by the positive stationary ion cores which can attract other electrons toward its direction of motion. The existence of such an attraction in the direction of motion can become instrumental in electron pairing if the same were true in a quantum mechanical treatment. Equation 2 has been redrawn in Fig. 4 for an electron speed equal to 0.01 $c$. The lobes in the direction of motion which have been designated with a positive sign represent the strength of the attraction potential while the portions which have been designated by negative sign represent the repulsion potential. (This speed was chosen to represent a typical material such as YBCO, which has a typical Fermi speed[3] around $10^7$ cm/sec and a dielectric constant around 100)

So far we have discussed a one dimensional lattice in which electrons are moving in one direction. In a three dimensional solid if electrons are allowed to move in all directions without a preferred direction of motion, the repulsion potential of one (area with negative sign in Fig. 4) will coincide with the attraction potential of another (area with positive sign in Fig. 4) and the effect of attraction potential will be averaged out. In other words, if the density of states around the Fermi surface is isotropic, no net attraction force will be experienced by electrons. On the other hand, if there is anisotropy in the density of states, a net attraction force will be developed in direction in which the density of states is larger. The optimum density of states would be one with the angular dependence similar to the attraction potential designated by positive sign in Fig. 4.

TOWARD A QUANTUM MECHANICAL DEVELOPMENT

Implicit in any quantum mechanical development of free electron behavior inside crystals is the charge neutrality. There are as many electrons as there are protons. If there is a charge imbalance, the system tries to change to its charge neutral state by the extra electric fields which are developed. An imbalance of charge causes either instability or a large affinity to attract positive or negative charges by the material. This is how we are going to attack this problem quantum mechanically. We assume a charge neutral crystal and add the charge imbalance due to relativity through a perturbation potential and calculate the scattering amplitudes that this potential causes. Since we are using a potential which is relativistic in nature, we should start with a relativistic quantum mechanical treatment of the problem. The Dirac equation for relativistic quantum mechanics can be adopted [6]. Solving the 4-dimensional space coefficients for our problem inside a crystal is not trivial but can be simplified by a justifiable approximation. Normally a particle is considered in non-relativistic regime when its energy is much less than its rest mass energy (or alternatively when its speed is much less than the speed of light). Considering that the energy of electrons at the Fermi level of regular crystals such as high temperature superconductors is much less than the electron rest energy which is around 511 kev a non-relativistic formulation of Dirac equation could be adopted. Dirac equation does not reduce to the Schrodinger equation in its non-relativistic approximation [6]. In fact a spin-magnetic field relation remains which results in spin-orbit coupling when the problem of an electron in the potential of a proton is considered. The interesting point is that although the wave functions can be approximated as non-relativistic, it is of great theoretical and practical importance to keep even the minute effect of spin-orbit coupling because it removes degeneracy through selection rules. Such interesting behavior regarding relativity is not peculiar to quantum mechanics. In the example of the attraction force between two current carrying wires similar situation occurs. The speed of electrons inside the wires is much less than the speed of light and we are safe to assume their speed as non-relativistic and apply regular Newtonian relations to their motion; however, the attraction force which is a purely relativistic affect does occur. The reason is of course the incredibly large number of electrons and protons inside the wires whose small relativistic affects add up to a measurable force.

We are going to argue in the same manner in this case. We are suggesting that the Dirac formulation of relativistic wave functions can be approximated by the non-relativistic wave functions in crystals in which the Fermi energy is much less than the electron rest energy. But although the wave functions can be considered as non-relativistic the change in charge neutrality as presented in this paper may have a noticeable effect on overall behavior of the electrons because although this change is very small the number of states which are involved in scattering due to this change, is so large that an appreciable superconducting gap may be produced. We are therefore adopting the non-relativistic free electron wave functions and band diagram inside crystals and adding equation 2 as a perturbation to their ground state. This is similar to adding spin-orbit coupling to non-relativistic quantum mechanics or speaking of a magnetic attraction between current carrying wires even though the moving electrons can be considered as non-relativistic for all other practical purposes.

Another point may be raised here that whether it is valid to apply a classical potential to wave like electrons in a quantum mechanical description. In other words, is it accurate to choose equation 2 as the perturbing potential? Whenever we have a classical potential of importance it might (and almost always) has an affect in quantum mechanics, granted that its affect is different from classical treatment and is not predicted by classical arguments. To find how such a classical term affects quantum mechanics we normally insert the classical potential into Schrodinger equation (or any appropriate equation) and obtain the results. The answer to the Schrodinger equation would tell us whether such a potential has a relevance in quantum mechanics or not, and will determine the response of the wave like electrons to such a classical term. It is in fact how quantum mechanical problems are dealt with. For example, in obtaining the affect of protons on electrons we just insert the classical Coulomb attraction into the Schrodinger equation, we do the same for repulsion potential between electrons in BCS theory, and we do the same for an electron in a harmonic oscillator potential. We do not need to define a new quantum potential unless the classical potential can be written in terms of quantum mechanical operators by the correspondence principle. Therefore, choosing equation 2 as the perturbing potential is valid and using it as a perturbing potential would tell us the response of the wave like electrons to such a classically defined potential.

TRANSITION APMPLITUDE AND THE GAP PARAMETER

To define the transition amplitude due to the attraction potential of equation 2 we will follow BCS theory as presented in Ref. 4. We treat equation 2 and the screened Coulomb repulsion as the perturbing potentials. The transition amplitude between initial pair states $(k^\uparrow, -k^\downarrow)$ and final pair states $(k'^\uparrow, -k'^\downarrow)$ is defined as:

$$V_{k,k'} = \langle k'^\uparrow, -k'^\downarrow | \Phi_S + \Phi | k^\uparrow, k^\downarrow \rangle, \qquad (3)$$

where $\Phi_S$ is the screened Coulomb repulsion between electrons and $\Phi$ is equation 2. Negative transition amplitude is needed for the occurrence of bound Cooper pairs. This occurs if the attraction potential is larger than the screened Coulomb repulsion and also if the initial and final states are chosen such that the repulsion part of $\Phi$ does not contribute as much as the attraction part to the integral.

The screened Coulomb potential is defined as [4,5]:

$$\Phi_S = \frac{q^2 \exp(-K_S r)}{r}, \qquad (4)$$

where $K_S$ is defined as $K_S^2 = 6\pi n q^2 / E_f$ in which $n$ is the electron density and $E_f$ is the Fermi energy. For a typical material such as YBCO with a density of electrons around $2 \times 10^{22}/cm^3$ [3], a Fermi energy of about 3.3 $ev$, and dielectric constant of 100, $K_S$ is about 0.1 A°. As a result the repulsion potential decreases to much less than $10^{-4} \times \frac{q^2}{r}$ in distances of 1 to 10 A°, which are the typical values for lattice constants. For the same material with an electron speed of $0.01c$ the strength of the attraction potential is equal to $10^{-4} \times \frac{-q^2}{r}$ in the direction of motion as given in Fig. 4. In other words, the attraction potential can become larger than screened potential for ordinary material. Keep in mind that the attraction potential of equation 2 is already a screened potential. When we say a crystal is charge neutral we are saying that the attraction potential of protons is screened or compensated by the electron potentials. The attraction potential of equation 2 is the minute change in such a screened potential which is equal to zero in all directions if electrons were stationary.

A larger attraction potential in some directions does not mean that the transition amplitude is negative for arbitrary initial and final scattering states. The reason is that attraction exists only around the direction of motion while in directions perpendicular to it repulsion is present. If initial and final scattering states are such that the attraction portion in Fig. 4 contributes to the integral (equation 3) more than the repulsion region contributes, then a negative transition is obtained. This could happen if the initial and final states have a symmetry similar to the attraction portion in Fig. 4. The above integral has been solved for selected states using MATLAB program. If we approximate $10^{-4} \times \frac{q^2}{r}$ as -1 $mev$ values of -4.5 $mev$, and -2.5 $mev$ are obtained for $V_{p,p} = \langle p|\Phi|p \rangle$, $V_{d,d} = \langle d_{z^2}|\Phi|d_{z^2} \rangle$, respectively. To understand why these wave functions result in negative transition amplitudes the integrals are shown in Fig. 5 pictorially. As can be observed electron probability density is mostly concentrated where the potential is attractive and in directions where repulsion is present the probability density is low. In this calculation we have considered the screened Coulomb repulsion to be negligible.

A negative transition amplitude means that electrons can condense into a superconducting state. Calculating the band gap of such a superconductor can follow the same steps as outlined in the BCS theory. In BCS only electrons in a shell of $\hbar\omega_D$ around Fermi energy [1,4] (where $\hbar\omega_D$ is the Debye energy) can contribute to lowering of energy.

Here we do not have this restriction and all the states can contribute to energy lowering if they have the required symmetry.

As was mentioned if the density of sates is isotropic in all directions, the attraction and repulsion contributions would average out. There must be a level of asymmetry in density of states for an energy lowering to occur. If we denote the density of states as a function which is both dependent on direction and energy, the following term can be assumed as a measure of how much anisotropy can contribute to energy lowering:

$$\alpha = \int_\Omega N_{(\Omega)} \Phi d\Omega \qquad (5)$$

in which $N_{(\Omega)}$ is the angular dependent part of the density of states. By definition if $N_{(\Omega)}$ is spherically symmetric, $\alpha = 0$, and if $N_{(\Omega)}$ has the same symmetry as $\Phi$, $\alpha = 1$. Even a slight asymmetry in density of states results in a value larger than zero. The value of this integral is multiplied by the regular density of states $N_{(E)}$ to obtain the gap parameter. We follow the same steps in BCS theory in obtaining the gap parameter. Assuming a constant density of states $N_{(E)}$ around the Fermi level (which in our case may not be appropriate because the energy range around Fermi energy is not bound by $\hbar\omega_D$), a constant value for $V_{k,k'}$, and $\alpha = 1$ we have [4]:

$$\frac{2}{N_{(E_f)} V_{k,k'}} = \int_{-\nabla E}^{+\nabla E} \frac{dE}{\sqrt{E^2 + \Delta^2}}, \qquad (6)$$

where the energies are written with respect to Fermi energy. The band gap is then obtained from:

$$\Delta = \frac{\nabla E}{\sinh[\dfrac{1}{N_{(E_f)} V_{k,k'}}]}. \qquad (7)$$

This equation is the same as BCS except that the energy runs over all energy levels which can result in a negative $V_{k,k'}$ and is not bound by $\hbar\omega_D$; therefore, for conventional density of states this integral can become larger than regular low Tc superconducting gaps. If we use the linear density of states which is given in reference 7 and replace it by a constant average, and assume d-type wave functions are involved in scattering, a gap of 50 *mev* for an energy interval of 15 *mev* is obtained.

A superconductor which is formed by this phenomenon is very much a BCS type superconductor except for its attraction potential which is different from the phonon-induced attraction.

CONCLUSIONS

The relativistic asymmetry in Coulomb repulsion potential leads to an attraction potential between electrons. The attraction potential is caused by uncompensated positive ion cores. We use this as a perturbation in Schrodinger equation to find its corresponding effects in quantum mechanics. It is observed that scattering amplitudes between p and d-wave functions become negative, which is the prerequisite for a Cooper pair bound state. To obtain a total lowering of energy in a three dimensional solid, the density of states must be anisotropic around the Fermi surface. There is no limit on the energy of the states which can contribute to the lowering of energy; as a result, even the small relativistic attraction potential can result in a relatively large superconducting gap parameter.

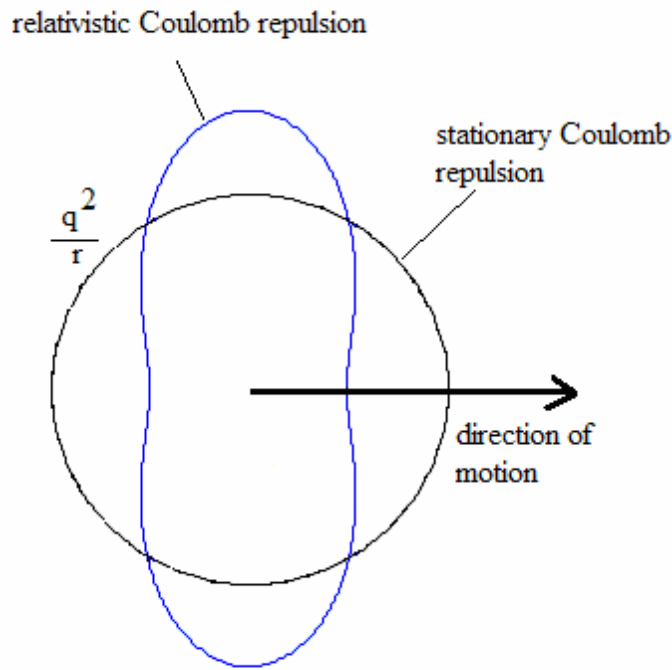

Fig. 1. For an stationary electron the Coulomb repulsion force is the circle which has a radius of $\frac{q^2}{r}$. For a moving electron with a speed of $0.7\,c$ the Coulomb repulsion changes as shown. It decreases in the direction of motion and increases in directions perpendicular to it. To obtain a three dimensional picture this shape must be rotated around the direction of motion.

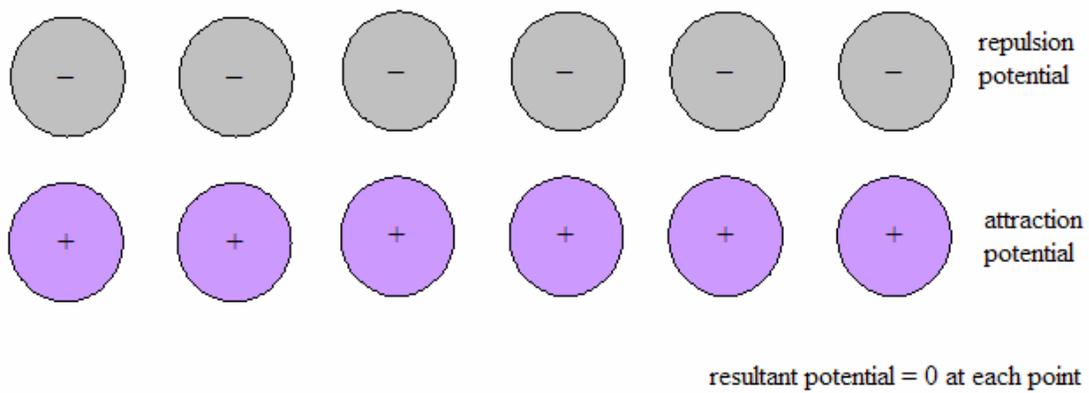

Fig. 2. Here a one dimensional crystal is shown in which for each positive ion there is an electron. The attraction potential of the cores is compensated by the repulsion potential of the electrons. The attraction and repulsion potentials are superimposed on top of each other but here they have been drawn separately for clarity.

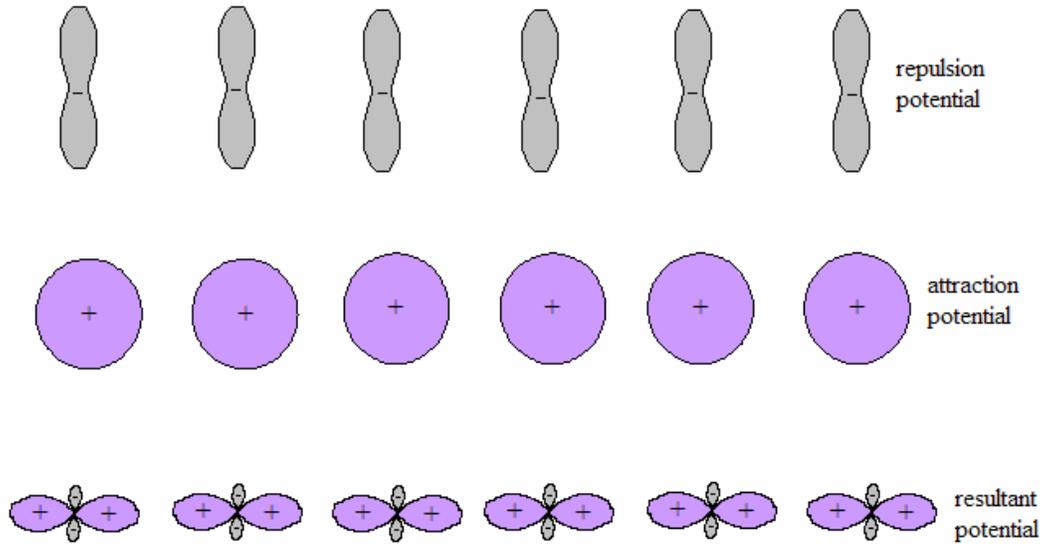

Fig. 3. If electrons are in motion their respective repulsion potential changes due to relativistic effects. This is shown pictorially. The spherically symmetric attraction potential of the ion cores is no longer compensated by this repulsion potential in all directions, a resultant potential is obtained which is attractive in the direction of motion and repulsive in directions perpendicular to it.

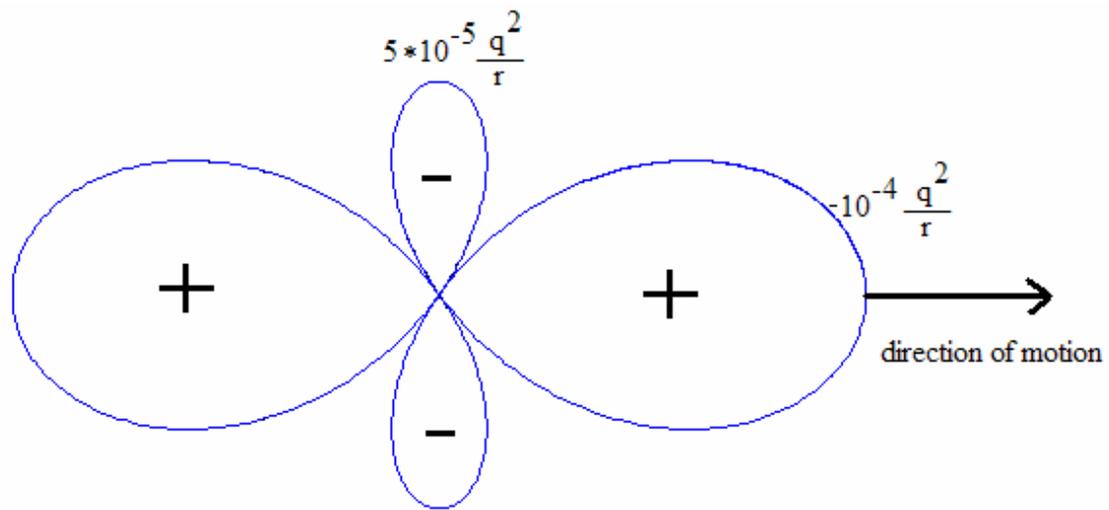

Fig. 4. The resultant potential when an electron is moving with a speed equal to 0.01 $c$.

Electrons will see an attraction force which is $10^{-4} \dfrac{q^2}{r}$ in the direction of motion.

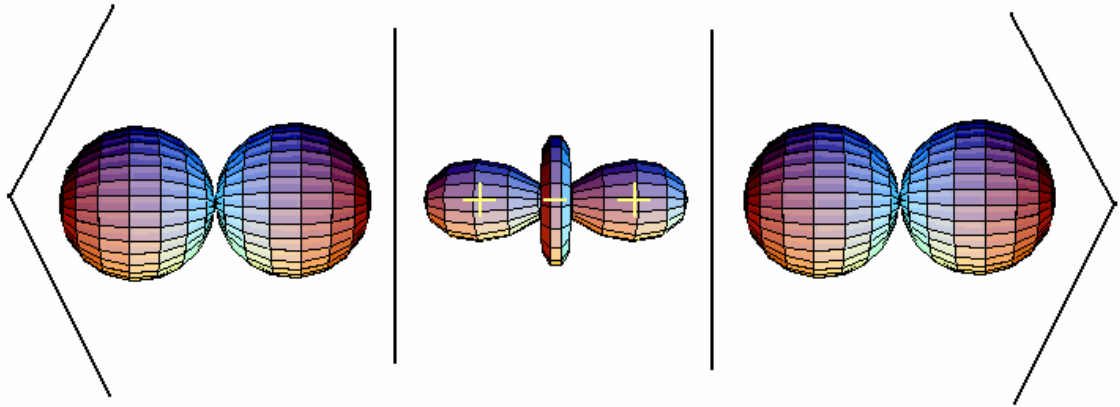

(a)

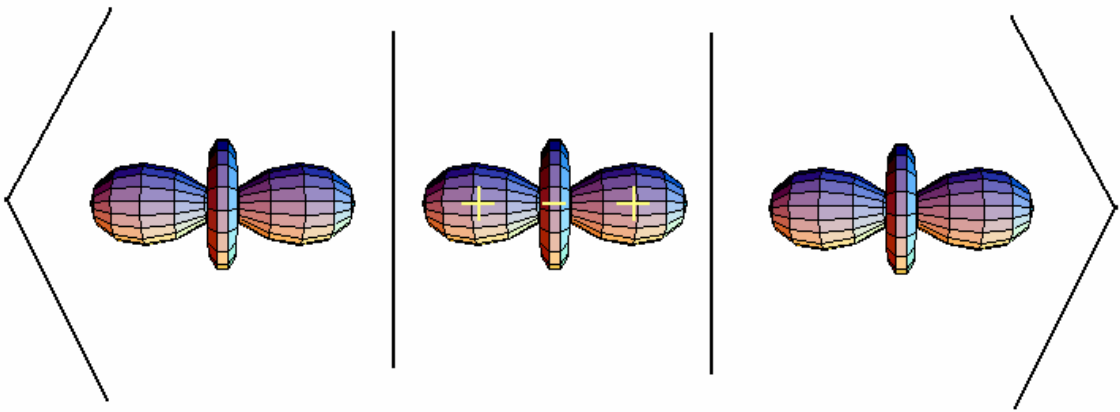

(b)

Fig. 5. The transition amplitude for p (a) and $d_{z^2}$ (b) states are shown pictorially ($d_{z^2}$ states are similar to $\Phi$ in their shape). As is observed such wave functions have their largest probability distribution in directions where attraction potential is present.